\input{aipcheck}

\documentclass[
    ,final            
  ]
  {aipproc}

\layoutstyle{6x9}

\begin{document}

\title{Quantization of Reissner-Nordstr\"{o}m Black Holes and Their Non-Singular Quantum Behavior}


\author{Mojtaba Taslimi Tehrani }{
  address={Department of Physics, Stockholm University, 10691 Stockholm, Sweden}
}

\author{Hoshang Heydari}{
  address={Department of Physics, Stockholm University, 10691 Stockholm, Sweden}
}

\begin{abstract}
Quantization of different regions of the  Reissner-Nordstr\"{o}m space time (charged black hole) is done in the framework of loop quantum gravity. The geometry of Reissner-Nordstr\"{o}m space-time is expressed in terms of Ashtekar variables which form the classical phase space of such a black hole. Using the loop quantization of phase space, the issue of singularity avoidance of such a black hole is addressed; based on spherically symmetry reduced models of loop quantum gravity, the operator analogue of the diverging factor of scalar curvature of the charged black hole is constructed and is shown to exhibit an upper bounded spectrum. This local criterion, together with the global one (non-singular quantum evolution equation) proves the avoidance of charged black hole singularity in symmetry reduced models of loop quantum gravity.
\end{abstract}

\maketitle

\section{Introduction}

Loop quantum gravity (LQG) is one of the active approaches toward constructing a quantum description of gravitational field. While currently there is no direct experimental test, presenting a consistent quantum description of unphysical behavior of classical gravity can serve as an internal test for such a tentative theory. One of such situations where classical general relativity fails to give a meaningful description happens at singularity of a black hole where a certain scalar curvature diverges. In this paper, we address the avoidance of this classical infinity for a charged black hole within the formalism of LQG \cite{Rovelli, Thiemann}.
To that end, we study the quantum behavior of a Reissner-Nordstr\"{o}m black hole by employing the spherically symmetric quantum geometry methods \cite{Bojowald,Bojowald2} to quantize the interior portion of such a space-time within which the classical singularity occurs.


\section{Classical Reissner-Nordstr\"{o}m Black Hole}

A spherically symmetric solution to Einstein-Maxwell equations is the Reissner-Nordstr\"{o}m metric which describes the space-time of a source with mass $M$ and charge $Q$  (in coordinate system $(x, \theta, \phi)$):
\begin{equation}\label{line element}
ds^2=-\left(1-\frac{2M}{x} + \frac{Q^2}{x^2}\right)dt^2+\left(1-\frac{2M}{x}+\frac{Q^2}{x^2}\right)^{-1}dx^2+x^2 d\Omega ^2.
\end{equation}
Horizons appear where $g_{00}=0$:
\begin{equation}
x^2-2Mx+Q^2=0 \longrightarrow x_{\pm}=M \pm \sqrt{M^2-Q^2}.
\end{equation}
For the  case of $M^2>x^2$ which is physically interesting, event horizons partition space-time into 3 regions: I $(x>x_+)$, II $(x_- <x<x_+)$, and III $(0<x<x_-)$. We observe that in region II, $x$ and $t$ interchange their roles and becomes time-like and space-like respectively, which defines a Kantowski-Sachs type homogeneous space. Region I and III, on the other hand, carry spherical symmetry.

\subsection{Classical Phase Space Variables}

\subsubsection{Region II}\label{II}

First we consider the region II in which the metric of space-time takes the form
\begin{equation}
ds^2=-\left(\frac{2m}{t}-\frac{Q^2}{t^2}-1\right)^{-1}dt^2 + \left(\frac{2m}{t}-\frac{Q^2}{t^2}-1\right)dx^2+ t^2 d\Omega ^2.
\end{equation}
According to definition of tetrad (frame) fields,
$
g_{\mu \nu}= \eta_{IJ} e^I_{\mu} e^J_{\nu},
$
they can be determined only up to a Lorentz transformation. This leaves us with an $SO(3,1)$ freedom in choosing tetrad. In fact, given the metric (\ref{line element}) we are free to choose their sign and Minkowski indices, which can be viewed as sort of a labeling 4 tetrad fields. 
The suitable choice for labeling 4 orthogonal frame fields turns out to be:
\begin{eqnarray}\nonumber \label{10}
e^0=\pm\left(\frac{2m}{t}-\frac{Q^2}{t^2}-1\right)^{-1/2}dt \textrm{ ; } e^1=\pm t sin\theta d\phi \\
e^2=\pm t d\theta \textrm{ ; } e^3=\pm \left(\frac{2m}{t}-\frac{Q^2}{t^2}-1\right)^{1/2}dx,
\end{eqnarray}
We constructed the $A$ field using spin connections compatible with \ref{10}:
\begin{equation}
A^3= \mp\gamma \left(\frac{Q^2}{t^3}-\frac{m}{t^2}\right) dx,
~~
A^2 = \pm \gamma  \left(\frac{2m}{t}-\frac{Q^2}{t^2}-1\right)^{1/2}d \theta,
\end{equation}
\begin{equation}
A^1 = \pm \gamma \left(\frac{2m}{t}-\frac{Q^2}{t^2}-1\right)^{1/2}sin\theta d\phi,
~~
A^3= \pm cos \theta d\phi.
\end{equation}
Moreover, the $E$ fields are given by:
\begin{equation}\nonumber
E_1 = \pm t \left(\frac{2m}{t}-\frac{Q^2}{t^2}-1\right)^{1/2} \partial_\phi,
~~
 E_2 =\pm t \left(\frac{2m}{t}-\frac{Q^2}{t^2}-1\right)^{1/2} sin\theta \partial_\theta,
\end{equation}
\begin{equation} \nonumber\label{Ex}
E_3 =\pm t^2 sin\theta \partial_x .
\end{equation}
If we had chosen other Minkowski indices for tetrad (\ref{10}) then we would not have obtained the conjugate pair $(A,E)$ with correct indices satisfying $\{A^i_a(x) , E^b_j(x') \}= \delta^i_j \delta^b_a \delta(x-x')$.
The phase space variables are determined up to a sign freedom. By demanding $E$ and $A$ to satisfy the diffeomorphism, Gauss and Hamiltonian constraints, their signs can be fixed relative to each other. The non zero components of diffeomorphism and Gauss constraints turn out to be:
\begin{equation}
H_\theta=\gamma t \left(\frac{2m}{t}-\frac{Q^2}{t^2}-1\right) cos \theta \left\{sgn(A^1_\phi)+sgn(A^2_\theta A^3_\phi)\right\},
\end{equation}
\begin{equation}
G_2= t \left(\frac{2m}{t}-\frac{Q^2}{t^2}-1\right)^{1/2}\cos \theta \left\{sgn(E^\theta_2 )+ sgn( A ^3_\phi E^{\phi1})\right\},
\end{equation}
and Hamiltonian constraint gives:
\begin{equation}
C=  t\left(\frac{2m}{t}-\frac{Q^2}{t^2}-1\right) sin^2\theta \left\{sgn(E^\theta_2) + sgn(E_1^\phi )\right\}.
\end{equation}
For the above constraints to be zero we must have:
\begin{equation}
sgn(E^\theta_2) = - sgn(E_1^\phi ),
~~
sgn(A^3_\phi) = +1,
~~
sgn(A^1_\phi) = - sgn(A^2_\theta ).
\end{equation}
This leaves us with two alternatives corresponding to the residual gauge freedom $(A_2, E_2) \rightarrow (-A_2, -E_2)$.
\subsubsection{Region I and III}
Next we consider the regions I and III.  The analogous calculations for these regions  with line element \ref{line element} leads to the following phase space coordinates
\begin{equation} \label{Ex1}
\textrm{I} \left\{
\begin{array}{l l}
\tilde{A}_a ^i= \tilde{c} \tau_{3} dr + \tilde{b}\tau_{2} d\theta +( cos\theta \tau_3 - \tilde{b} sin\theta \tau_1 )d\phi\\
\tilde{E}_i ^a= \tilde{p}_c \tau_3 sin\theta \partial_{r} +\tilde{p}_b \tau_2 sin \theta \partial_{\theta}-\tilde{p}_b \tau_1 \partial_ {\phi},\\
\end{array} \right.
\end{equation}
\begin{equation}  \label{Ex2}
\textrm{II } \left\{
\begin{array}{l l}
\tilde{A}_a ^i= \tilde{c} \tau_{3} dr - \tilde{b}\tau_{2} d\theta +( cos\theta \tau_3 +\tilde{b} sin\theta \tau_1 )d\phi\\
\tilde{E}_i ^a= \tilde{p}_c \tau_3 sin\theta \partial_{r} - \tilde{p}_b \tau_2 sin \theta \partial_{\theta}+ \tilde{p}_b \tau_1 \partial_ {\phi},\\
\end{array} \right.
\end{equation}
where 
$
\tilde{b}=\pm \gamma \left(1-\frac{2m}{x}+\frac{Q^2}{x^2}\right)^{1/2} \textrm{ ; } \tilde{c}=\mp \gamma \left(\frac{m}{x^2}-\frac{Q^2}{x^3}\right)
$ $
\tilde{p}_c = \pm x^2 \textrm{ ; } \tilde{p}_b =x\left(1-\frac{2m}{x}+\frac{Q^2}{x^2}\right)^{1/2}.
$
This corresponds to variables introduces in \cite{Bojowald,Bojowald2} as
\begin{equation}
A_x=\tilde{c} \textrm{ , } E^x = \tilde{p}_c \textrm{ ; }
~~
\gamma K_\phi= \tilde{b} \textrm{ , } E^\phi=\tilde{p}_b \textrm{ ; }
~~
\eta=(2n+1) \pi \textrm{ , } P^\eta=0,
\end{equation}
which constitute a 4 dimensional phase space.

The singularity occurs at $x=0$, or equivalently $E^x=0$, where the scalar curvature become infinite and lies in the spherically symmetric region III.

\section{Curvature Boundedness and Singularity avoidance }
In this section, we make use of the crucial points regarding the foundations and results of spherically symmetric quantum geometry developed by Bojowald \cite{Bojowald,Bojowald2}. The reader is referred to the original articles for many technical subtleties which are omitted inevitably.
The  Kretschmann scalar curvature of the Reissner-Nordstr\"{o}m black hole is defined by \cite{Kretschmann}
\begin{equation} \label{RNcurvature}
 R^{\mu \nu \rho \sigma} R_{\mu \nu \rho \sigma}= \frac{48 M^2 x^2 - 96 M Q^2 x + 56 Q^2 }{x^{8}},
\end{equation}
From this expression, we could observe that the irremovable curvature singularity of a charged black hole occurs at $x=0$. It lies in the spherically symmetric region III of the Reissner-Nordstr\"{o}m space-time. We will show that the quantum operator corresponding to $\frac{1}{x}$, on the other hand,  exhibits a spectrum which is bounded above. To construct such an operator, we need the classical function to be expressed suitably in terms of well-defined operators in the reduced theory.  A candidate to serve as the desired classical function is $\frac{sgn(E^x)}{ \sqrt{|E^x(x)|}}=\frac{1}{x}$. However, we cannot naively replace the inverse squired of triad with its operator analogue since it has zero eigenvalue. But, as a Poisson bracket of functions having well-defined operators, its corresponding quantum operator can be realized by replacing Poisson bracket with ($i \hbar$ times) commutator.
This is done by the following relation:
\begin{equation}
\mathcal{R} \equiv \frac{sgn(E^x)}{ \sqrt{|E^x(x)|}} = \frac{1}{2 \pi \gamma G}  \left\{A_x(x),  \sqrt{|E^x(x)|}\right\}.
\end{equation}
$\sqrt{|\hat{E}^x(x)|}$ can be defined as a diagonal operator whose eigenvalues are square roots of (absolute value of) those of  $\hat{E}^x(x)$, since $\hat{E}^x(x)$ is diagonal on spin network states.

As is usual in LQG, we express $\mathcal{R}$  to the order of $\epsilon^2$ as:
$$
\mathcal{R} = \frac{1}{2 \pi \gamma G}  tr \left( \tau_3 h_x\{h_x^{-1}, \sqrt{|E^x(x)|}\}\right),
$$
where along the inhomogeneous direction $x$, we can expand holonomy $h_x(x)$ as:
\begin{equation}
h_x=exp\left(i\int _e dx A_x(x)\right) \approx  1 + i\epsilon A_x(\nu),
\end{equation}
with $\epsilon=\nu_+ - \nu$ being the coordinate distance between two vertices $\nu$ and $\nu_+$ connected by the edge e.
And consequently its corresponding operator as:
\begin{eqnarray}
\hat{\mathcal{R}} &= &\frac{1}{2 \pi \gamma \ell_{Pl}^2}  tr \left( \tau_3 \hat{h}_x \left[ \hat{h}_x^{-1}, \sqrt{|\hat{E}^x(x)|} \right]\right)
\\\nonumber&
 =&\frac{1}{2 \pi \gamma \ell_{Pl}^2} \left( cos(\frac{1}{2} \int A_x)\sqrt{|\hat{E}^x(x)|} sin(\frac{1}{2} \int A_x) -  sin(\frac{1}{2} \int A_x)\sqrt{|\hat{E}^x(x)|} cos(\frac{1}{2} \int A_x)  \right),
\end{eqnarray}
where we have used $h_x=exp\left(i\int _e dx A_x(x)\right)=cos(\frac{1}{2} \int A_x) + 2 \tau_3  sin(\frac{1}{2} \int A_x).$

The kinematical Hilbert space of the  the spherically symmetric reduction of LQG is the space spanned by spin network state $T_{g, k, \mu}$ which satisfy the Gauss constraint:
\begin{eqnarray}
T_{g, k, \mu}&=& \prod_{e \in g} exp\left(\frac{i}{2}k_e \int _e dx ( A_x(x)+\eta^{\prime})\right) \prod_{\nu \in V(g) } exp \left(i \mu_\nu \gamma K_\phi (\nu)\right).
\end{eqnarray}
For a  graph $g$, these are cylindrical functions of holonomies along edges $e$ of $g$. Edges are labeled by irreducible representations of $U(1)$, while vertices $V(g)$ of such spin networks are labeled by irreducible $\bar{\mathbf{R}}_{Bohr}$ representations $\mu_\nu \in \mathbf{R}$ and irreducible $S^1$ representation $k_\nu \in \mathbf{Z}$.
The action $\hat{\mathcal{R}}$  on spin network states
 gives rise to the spectrum
\begin{eqnarray} \nonumber
\hat{\mathcal{R}} T_{g, k, \mu} &=& \frac{1}{2  \pi \sqrt{\gamma} \ell_{Pl}} \left(\sqrt{\frac{1}{2} |k_{e^+(x)}+k_{e^-(x)}+2|}-\sqrt{\frac{1}{2} |k_{e^+(x)}+k_{e^-(x)}-2|} \right)T_{g, k, \mu}.
\end{eqnarray}
Such an operator  is diagonal on spin network states and therefore commutes with all operators in kinematical Hilbert space. Moreover, it has a bounded spectrum with maximum value of $ \frac{1}{\sqrt{2} \pi \sqrt{\gamma} \ell_{Pl}}$. Thus, the scalar curvature \ref{RNcurvature} at quantum level has a maximum value of \cite{Mojtaba}:
\begin{eqnarray}  \label{RNcurv}
\left( R^{\mu \nu \rho \sigma} R_{\mu \nu \rho \sigma}\right)_{max}&=& \frac{1}{\gamma^3 \pi^6 \ell^{6}_{Pl}} \left( 6 M^2  - \frac{96  M Q^2}{\sqrt{2} \pi \sqrt{\gamma} \ell_{Pl}}+ \frac{28 Q^2}{ \pi^2 \gamma \ell^2_{Pl}}\right).
\end{eqnarray}
This reflects the fact that the divergence of $\frac{1}{x}$ is kinematically well-behaved at quantum level.
In this paper, we have built the classically divergent Kretschmann singularity out of Ashtekar variables for the spherically symmetric part of a Reissner-Nordstr\"{o}m black hole. We used the methods of spherically symmetric reduction of LQG, and constructed the operator analogue of such a curvature. This construction shows that quantum mechanically the scalar curvature is bounded above which indicates local avoidance of singularity in such a symmetry reduced model. However, the true cure of the singularity must be done by studying the dynamics which is made in
 \cite{Bojowald2}.

\begin{theacknowledgments}
 The  work was supported  by the Swedish Research Council (VR).
\end{theacknowledgments}



\bibliographystyle{aipproc}   




\end{document}